\documentclass[conference]{IEEEtran}
\IEEEoverridecommandlockouts
\usepackage{cite}
\usepackage{amsmath,amssymb,amsfonts}
\usepackage{balance}
\usepackage{graphicx}
\usepackage{textcomp}
\usepackage{xcolor}
\usepackage{epstopdf}
\def\BibTeX{{\rm B\kern-.05em{\sc i\kern-.025em b}\kern-.08em
    T\kern-.1667em\lower.7ex\hbox{E}\kern-.125emX}}
\usepackage{booktabs}  
\usepackage[group-separator={,}]{siunitx}
\usepackage{csquotes}  
\usepackage{hyperref}
\usepackage{scalerel}
\usepackage{subfigure}
\usepackage[linesnumbered,ruled,vlined]{algorithm}
\usepackage[noend]{algpseudocode}

\usepackage{chemfig}
\usepackage[version=3]{mhchem}
\usepackage{datetime}
\usepackage{tikz}
\usetikzlibrary{shapes.geometric, arrows,chains}
\usetikzlibrary{decorations.pathmorphing} 
\usetikzlibrary{matrix} 
\usetikzlibrary{arrows} 
\usetikzlibrary{calc} 
\usetikzlibrary{svg.path}
\tikzstyle{block} = [draw,rectangle,thick,minimum height=2em,minimum width=2em]
\tikzstyle{sum} = [draw,circle,inner sep=0mm,minimum size=2mm]
\tikzstyle{connector} = [->,thick]
\tikzstyle{line} = [thick]
\tikzstyle{branch} = [circle,inner sep=0pt,minimum size=1mm,fill=black,draw=black]
\tikzstyle{guide} = []
\tikzstyle{snakeline} = [connector, decorate, decoration={pre length=0.2cm,
                         post length=0.2cm, snake, amplitude=.4mm,
                         segment length=2mm},thick, magenta, ->]

\usepackage[pscoord]{eso-pic}

\definecolor{orcidlogocol}{HTML}{A6CE39}
\tikzset{
  orcidlogo/.pic={
    \fill[orcidlogocol] svg{M256,128c0,70.7-57.3,128-128,128C57.3,256,0,198.7,0,128C0,57.3,57.3,0,128,0C198.7,0,256,57.3,256,128z};
    \fill[white] svg{M86.3,186.2H70.9V79.1h15.4v48.4V186.2z}
                 svg{M108.9,79.1h41.6c39.6,0,57,28.3,57,53.6c0,27.5-21.5,53.6-56.8,53.6h-41.8V79.1z M124.3,172.4h24.5c34.9,0,42.9-26.5,42.9-39.7c0-21.5-13.7-39.7-43.7-39.7h-23.7V172.4z}
                 svg{M88.7,56.8c0,5.5-4.5,10.1-10.1,10.1c-5.6,0-10.1-4.6-10.1-10.1c0-5.6,4.5-10.1,10.1-10.1C84.2,46.7,88.7,51.3,88.7,56.8z};
  }
}

\newcommand\orcidicon[1]{\href{https://orcid.org/#1}{\mbox{\scalerel*{
\begin{tikzpicture}[yscale=-1,transform shape]
\pic{orcidlogo};
\end{tikzpicture}
}{|}}}}

\begin{document}

\title{A Reinforcement Learning Approach for Performance-aware Reduction in Power Consumption of Data Center Compute Nodes}
\author{\IEEEauthorblockN{Akhilesh Raj}
\IEEEauthorblockA{\textit{Department of Electrical Engineering} \\
\textit{Vanderbilt University}\\
Nashville, Tennessee, USA\\
akhilesh.raj@vanderbilt.edu \orcidicon{0000-0001-6639-7432}}
\and
\IEEEauthorblockN{Swann  Perarnau}
\IEEEauthorblockA{\textit{Mathematics and Computer Science Division} \\
\textit{Argonne National Laboratory}\\
Lemont, Illinois, USA\\
swann@anl.gov \orcidicon{0000-0002-1029-0684}}
\and
\IEEEauthorblockN{Aniruddha Gokhale}
\IEEEauthorblockA{\textit{Department of Computer Science} \\
\textit{Vanderbilt University}\\
Nashville, Tennessee, USA \\
a.gokhale@vanderbilt.edu \orcidicon{0000-0002-7706-7102}}
\and
}

\maketitle

\begin{abstract}
As Exascale computing becomes a reality, the energy needs of compute nodes in cloud data centers will continue to grow. 
A common approach to reducing 
this energy demand is to limit the power consumption of hardware components when workloads are experiencing bottlenecks elsewhere
in the system. However, designing a resource controller capable of detecting and limiting power consumption on-the-fly is a complex issue and can also adversely impact application performance. In
this paper, we explore the use of Reinforcement Learning (RL) to design a power capping policy on cloud compute nodes using observations
on current power consumption and instantaneous application performance (heartbeats).
By leveraging the Argo Node Resource Management (NRM) software stack in conjunction with the Intel Running Average Power Limit (RAPL) hardware
control mechanism, we design an agent to control the maximum supplied power to processors without compromising on application
performance. Employing a Proximal Policy Optimization (PPO) agent to learn an optimal policy on a mathematical model of the compute
nodes, we demonstrate and evaluate using the STREAM benchmark how a trained agent running on actual hardware can take actions by balancing power consumption and application
performance.
\end{abstract}
            

\section{Introduction} \label{sec:Introduction}
Cloud data centers are increasingly providing a variety of high performance computing nodes 
including hardware accelerators to support the needs of a range of compute-intensive 
applications including various machine learning training/inferencing and scientific
applications. In these setups, a large number of compute nodes are networked together via
high-speed networks to provide high performance computing capabilities. 
This growth, however, has given rise to several challenges, notably in the areas of energy efficiency.  
For instance, Darrow et al. in ~\cite{darrow2009opportunities} pointed out that data centers consume 10-50 times the energy per floor space compared to any other commercial building and the consumed energy constitute 2\% of the total US energy consumption.
Multiple surveys and studies~\cite{darrow2009opportunities,koomey2011growth,koot2021usage,shalf2020future,dayarathna2015data} have shown that the global power consumption for computing is expected to increase in the upcoming years. 

Consequently, research on data center power consumption has gained considerable attention since the publication of initial statistics on its energy usage, which illuminated the rapidly increasing demands of servers worldwide~\cite{koomey2011growth}.
A more recent survey shows that the growing demands of these data centers are not in proportion with their efficiency despite the technological innovations ~\cite{koot2021usage}. In other words, the rate at which the power consumption increases across the generations of data centers are higher than the rate of growth in their efficiency. 
Within the data center, almost 86\% of the total power consumption is equally shared between the cooling system and servers~\cite{shehabi2016united}, which makes the power consumption of compute nodes an important issue that needs to be addressed. 

In the context of reducing compute node power consumption, various methods have been proposed that can be broadly classified into two categories: those that focus on minimizing power dissipation and those that focus on minimizing power supply. These approaches aim to reduce the overall energy consumption of compute nodes, while maintaining their performance and functionality.
The former is strictly a Very Large Scale Integration (VLSI) design problem, while the latter is a cyber-physical systems problem, which can be dealt with using efficient power management algorithms (i.e., control system design), which is the focus of this paper. 

Specifically, we design a method to reduce the consumed power without any noticeable impact on the execution time of a workload. Through our work, we achieve regulation of the average supplied power using a reinforcement learning (RL) agent, trained using a mathematical model which relates the progress made by the application towards the completion of its scientific-goal to the power cap (pcap) of RAPL actuators. Additionally, the goal is to not let the execution time vary significantly when compared to that with maximum power execution.

To that end, we present a real-time, dynamic power management scheme that is based on observing both time-aware and power-aware variables as a solution to the power management problem of the high performance compute nodes. 
In our work, the RL agent observes the progress made by the application towards completing its scientifc goal at each instant and makes a decision on the power requirement for the next time step based on this observation. The action, which is based on this decision, is taken by setting the running average power limit (RAPL) power caps at the computed value. RAPL is an interface available on modern Intel processors to monitor and control the energy consumption of various power domains. Our approach is incorporated in the Argo Node Resource Management (NRM) stack~\cite{anrm}.


The key contributions of this paper are the following:
\begin{itemize}
\item A novel RL-based algorithm that relies on the system model for training.
\item An RL-based architecture employing the trained model to control the compute node within a desired operational region, while utilizing real-time progress measurements.
\item An implementation of the algorithm with an open-source repository for researchers to test and extend the results.
\end{itemize}

The remainder of this paper is structured as follows: In Section ~\ref{sec:Related_work}, we review existing algorithms in power and performance optimization. Section~\ref{sec:Background} presents the necessary background for this paper, including the RL algorithm, hardware and software stacks, and packages used in the experiments. Section ~\ref{sec:Methodology} outlines the proposed algorithm and workflow in the context of compute nodes. In Section ~\ref{sec:Eval}, we present the experimental design and implementation, along with the results and analysis. Finally, Section ~\ref{sec:Conclusion} provides conclusions including a brief summary, the impact of this work and future opportunities. 


\section{Related Work}\label{sec:Related_work}

{Prior methods for regulating the performance on HPC nodes can be classified either as a Power-aware scheme~\cite{hsu2005power,patki2015practical} or as a Time-aware scheme~\cite{sadrosadati2019itap,qiu2012three}. A power-aware scheme is designed with the primary objective of minimizing power consumption while still achieving acceptable performance levels. The main focus of this scheme is to optimize energy usage and reduce power consumption in computing systems. Power-aware schemes may employ techniques such as dynamic voltage and frequency scaling (DVFS), power capping, power gating, and other power management methods to control and regulate power usage based on workload and system conditions. The goal is to maximize energy efficiency and reduce the overall carbon footprint of the computing system. Time-Aware Scheme on the other hand, prioritizes meeting specific time or performance requirements, often associated with critical real-time or time-sensitive applications. The main objective of a time-aware scheme is to ensure that tasks or processes are completed within predefined time constraints or deadlines. In time-aware schemes, performance and response time take precedence over power efficiency. These schemes may use techniques like aggressive clock frequency scaling, parallelism, or task prioritization to meet the time requirements of critical applications.} This section provides an overview of some of these methods which also explains why there is a growing demand for machine learning based methods for the power and performance control that has lead to the proposed work. 

Our prior work~\cite{caglar2014iplace} researched software approaches for power optimization using duty cycle, voltage and frequency scaling, wherein we used machine learning techniques to model the application performance and hardware characteristics, and use these models to guide the placement decisions.
But, the choice of method highly depended on the hardware under consideration.
Similarly, Jung et al. \cite{jung2010supervised} proposes a supervised learning based power management for multi-core processors. However, such algorithms are not always reliable due to the need for training data prior to the implementation of software-based optimization. 

On the other hand, methods that deal with system architecture that set the input power based on the system requirements converges faster. For example, a RAPL-based scheme supports a variety of algorithms that relies on hardware-based control for power capping on Intel processors~\cite{david2010rapl}. But the automation/feedback required for the processor to determine the power requirements based on the application it runs is not detailed in the paper.
Similarly Raghavendra et al. ~\cite{raghavendra2008no} utilizes a controller to stop the supplied power that is being delivered to the subsystems, by detecting the no-load zones in the application during its execution. Utilizing the technique, they are able to regulate the power to finer granularities. But, the efficiency of detecting the workload characteristics is not at the best, since the research focuses only on the hardware centric approach which is the main contribution of the paper.

Zhang et al. ~\cite{zhang2016maximizing} later came up with a hybrid method, combining software and hardware based methods to control the power supply, which showed significant improvement over the other methods. The authors combined the efficiency of the software-based approaches like DVFS, Task Scheduling etc. and the timeliness of the hardware approaches like Power Gating, Voltage regulators etc. together by reducing the time it takes to implement a power cap on the RAPL sensor from the moment its value is set. But a workload-based adaptive control of power was not considered here. Petoumenos et al. ~\cite{petoumenos2015power}, through their research give a comparative analysis of the software, hardware and hybrid approaches towards the power optimization in compute nodes.

Cerf et al. in her recent research  ~\cite{cerf2021sustaining} proposed a control theoretical approach to understand the possible operating regions of a compute node at which the performance decay is minimized under a reduced power supply. A classical proportional-integral-derivative (PID) controller was designed to impose power cap (PCAP) for RAPL actuators on an Intel(R) Xeon(R) Gold 6126 CPU by using performance feedback from the compute nodes~\cite{ramesh2019understanding}.
The RAPL actuators comprising a control knob or the PCAP knob and a time-window knob were used to regulate the average power given between two sampling intervals to a compute node. A mathematical model formulated using static characterization was used for the design of the controller. The controller was shown to be working for the compute nodes by making it track a user-computed set-point for the desired performance. While an adaptive controller~\cite{hawila:hal-03765849} addressed model inaccuracies, reliance on a user-defined set point calls for an automated approach.

Considering the limitations in prior work, we emphasize the design of an optimal controller by solving the underlying power and performance optimization problem.  We propose a hybrid approach for determining the best operating power cap for a compute node by relying on the instantaneous performance analysis that is obtained using the state-of-the-art algorithms.

\section{Background}\label{sec:Background}
To make the paper self-contained, this section provides the necessary background on the concepts and technologies used.
\subsection{Reinforcement Learning} \label{subsec:Reinforcement_Learning}
Reinforcement Learning (RL) iteratively improves a sequence of actions or strategy through continuous interactions with the environment over a finite or infinite amount of time traversed using discrete time steps. 
In this work we use an RL agent to learn an optimal control policy by constantly generating an action and evaluating it using a mathematical model that describes the relation between power cap  and progress made by the compute node while running an application. The RL problem can be mathematically formulated as a Markov Decision Process (MDP), where the agent interacts with an environment that is modeled as a set of states, actions, and rewards. 
At each time step, the agent observes the current state, selects an action based on a policy, and receives a reward from the environment. The goal of the agent is to learn a policy that maximizes the expected cumulative reward over time. 
In this paper we use a model-based algorithm by employing a Proximal Policy Optimization (PPO)~\cite{schulman2017proximal} agent to learn the state transition and the reward functions, thereby learning the policy.

\subsection{Performance Measurement} \label{subsec:Performance_Measurement} Instantaneous measurement of the controlled variable is an important aspect of feedback control problems. The RL algorithm for maximizing performance under a controlled power cap being an optimization problem, also requires the measurement of instantaneous performance. During the measurement, care must be taken that it does not interfere with the overall working and thereby the performance of the system.  Therefore, we use a lightweight instrumentation library~\cite{ramesh2019understanding} and follow the instrumentation steps outlined in ~\cite{cerf2021sustaining}.  We formally re-define the progress measurement equation introduced in the paper \cite{ramesh2019understanding} considering the number of messages received between two sampling instants. Therefore in this paper the progress at the time instant $t_i$ is given by:
 \begin{equation}
    \begin{aligned}
                progress(t_i) = \underset{\forall k,t_k \in [t_{i-1},t_i]}{median}\bigg(\frac{N}{t_k-t_{k-1}} \bigg).
    \end{aligned}
    \label{eq:progress-calculation}
\end{equation}
where $N$ is the number of messages received between the time instants $t_k$ and $t_{k-1}$.


\subsection{Intel RAPL} \label{subsec:RAPL} The running average power limit (RAPL) is an interface available with modern Intel processors for power monitoring and controlling making it unavoidable in energy efficient computing. It also allows users to specify a power cap on available hardware domains using model-specific registers or the associated Linux sysfs subsystem. 
The RAPL interface uses two knobs, the power-limit knob and a time window knob to supply the average power for a user-defined period of time. The internal controller then guarantees that the average power over the time window is maintained. This mechanism also offers sensors through the same interface to measure the total energy consumed since the processor was turned on. Consequently, RAPL can be used for both measuring and limiting power usage~\cite{rountree2012beyond}.

\section{Methodology}\label{sec:Methodology}

This section describes details of our model-based reinforcement learning (RL) which relies on a mathematical model representing the relation between the progress and the power cap (PCAP) in a compute node for training the RL agent. 
In our previous work~\cite{cerf2021sustaining}, where the performance of a compute node under a given application was controlled using a suitable PID controller running alongside the application, itself utilized all the available cores of the Skylake processor. However, a similar approach is not possible in the current work for training the RL agent because of the time and computational complexity involved. Therefore, we rely on the mathematical model for training the RL agent. 

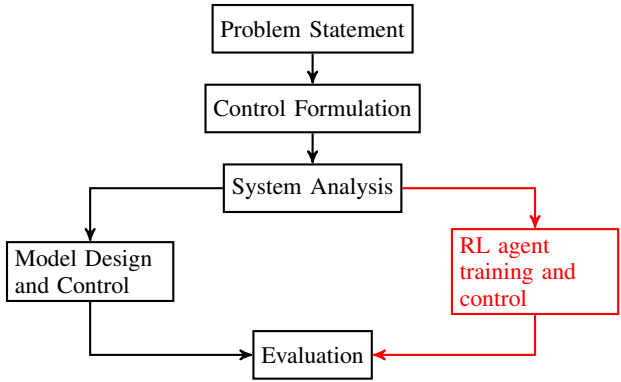
\begin{figure}[htb]
\centering
\begin{tikzpicture}[scale=1, auto, >=stealth']
\centering
\small
\matrix[ampersand replacement=\&, row sep=0.2cm, column sep=0.4cm] {
\& \&  \node[block](state_1){Problem Statement};\&  \& 
\\
\\
\& \& 
\node[block](state_2){Control Formulation};\& \& 
\\
\\
\& \& 
\node[block](state_3){System Analysis};\& \& \& 
\\
\&
\node[block](state_4)[text width=2cm]{Model Design and Control}; \& \& \node[block](state_5)[text width=2cm,color=red]{RL agent training and control}; \& \
\\
\& \&
\node[block](state_6){Evaluation}; \& \&
\\
};
\draw [connector] (state_1) -- node{}(state_2);
\draw [connector] (state_2) -- node{}(state_3);
\draw [connector] (state_3) -| node{}(state_4);
\draw [red] [connector] (state_3) -| node{}(state_5);
\draw [connector] (state_4) |- node{}(state_6);
\draw [red] [connector] (state_5) |- node{}(state_6);
\end{tikzpicture}
\vspace{-0.8cm}
\caption{A pictorial representation showing the differences in methodologies implementing a classical control and an RL based control. The RL method discussed in this paper is colored in red.}
\label{fig:block_diagram}
\vspace{-0.5cm}
\end{figure}

Figure~\ref{fig:block_diagram}  depicts our methodology while comparing it with the steps involved in the design of the PI controller in our prior work~\cite{cerf2021sustaining}. The processes involved in the two approaches beginning from the formulation of problem statement and extending till the evaluation are shown in the block diagram in Figure~\ref{fig:block_diagram}. It can be observed that the initial problem statement, the steps leading to the control formulation, and the system analysis are same in both the approaches. But, when it comes to the controller block, the PI controller relies on standard tuning methods from the classical control theory to determine its control gains, (i.e tuning the proportional, integral and derivative constants, $K_p, K_d \text{ and } K_i$). In contrast, the RL method uses an iterative, policy improvement and policy update process by using an actor-critic network. 
It is only in this stage of the controller design that the two algorithms take different routes in the implementation. The rest of this section details the steps used in our approach, i.e., the branch shown in red color. 

\subsection{Problem Definition}\label{subsec:Problem_statement}  
The first step is to define the objectives and the constraints involved in the optimization problem.
Our objective is to maximize performance within a given power cap (PCAP) constraint, while taking into account the relationship between progress and PCAP as determined by the node dynamics.
In other words, we aim to improve the execution time of the application while still using the minimum  allowed energy. A simple way of ensuring this behavior is by making sure that the computational progress is at its peak under the given PCAP.
\subsection{Mathematical modeling and MDP identification} \label{subsec:Mathematical_Modeling}
Given that the current progress of our problem is dependent on both the previous progress and the current power cap, it is expected that the node dynamics follow the characteristics of a Markov Decision Process (MDP). 
Thus, to effectively apply a reinforcement learning to a problem, it is crucial to identify the MDP tuple $(\mathcal{S}, \mathcal{A}, \mathcal{P},r,\gamma)$ representing the state space, a set of actions available to an agent, the unknown transition kernel, the reward function and the discount factor, respectively. In addition to identifying the MDP, we also need to derive a mathematical model that represents the node dynamics so as to train the RL agent. To support the claim that the node dynamics have MDP characteristics and to derive the mathematical model, we use a static characterization approach and utilize line-fitting algorithms from SciPy~\cite{2020SciPy-NMeth}.

The experiment, which is part of the model building process, begins by collecting data points in the form of input-output pairs, where the input (i.e., PCAP) is provided at certain pre-defined instances and the output (i.e., progress made by the compute node on the workload) is calculated using Equation~\eqref{eq:progress-calculation}. 
These data points correspond to an entire benchmark execution where a constant PCAP is applied, and the progress signal is averaged. To model the time-averaged relationship between power cap and progress, we consider only stabilized situations of the progress value generated using PCAPs allowed by the RAPL actuators. At least 20 experiments are run for each compute node to generate enough data points. Based on these experiments, a static model is obtained by linking the time-stabilized power cap to the progress using line-fitting algorithms. We observed that the PCAP-progress relation follows an exponential relationship, as shown by Equation~\eqref{non_linear_equivalent}:
\begin{equation}
    \begin{aligned}
          progress = & K_{L}(1-e^{-\alpha(a.PCAP+b-\beta)}),\\
    \end{aligned}
    \label{non_linear_equivalent}
\end{equation}
where $a$ and $b$ parameters represent RAPL actuator accuracy of the node (slope and offset, respectively). Therefore, the actual power applied on a node can be approximated as: $a.PCAP + b$ which is also the measured power. $\alpha$ and $\beta$ characterize the benchmark-dependent power-to-progress profile and $K_L$ is the linear gain and is both benchmark- and node-specific. The effective values of the fitting coefficients given by $\alpha, \beta, a, b, K_L$ and $ \tau $ are obtained using the above mentioned steps, where the line fitting problem follows a non-linear least squares optimization.
The linear characterization of dynamics can be then computed using least square optimization~\cite{cerf2021sustaining} and is given as
\begin{equation}
            \begin{aligned}
                progress_L (t_{i+1}) = & \frac{K_L \Delta t_i}{\Delta t_i+\tau}.PCAP_{L}(t_i) \\& +\frac{\tau}{\Delta t_i + \tau} progress_L(t_i),
            \end{aligned}
            \label{eq:linear_mathematical_model}
        \end{equation}
where
$\Delta t_i = t_{i+1} - t_i$ is the control interval and $\tau$ is the time constant. The values of $\Delta t_i$ and $\tau$ are chosen as $1 \unit{\second}$ and $\frac{1}{3}\unit{\second}$, respectively, for the experiments. We can use either of these models for training.

After defining the Markov Decision Process (MDP), we proceed to assign distinct characters to each role in the MDP, which include the state, action, and observation spaces that were previously established. In this study, the current state, denoted as $s^{'} \in \mathcal{S}$, is represented by the progress value at time $t$, $progress_t$. On the other hand, the action, denoted as $a \in \mathcal{A}$, taken by the agent, is determined by the $PCAP$ value at time $t$ based on the observation $progress_{t-1}$ ($s \in \mathcal{S}$), which represents the progress value at the time $t-1$.

\subsection{Reward function}\label{subsec:Reward_function}
We now explain how the reward function is chosen to facilitate fast and accurate learning.
The reward at a given time $t$ reflects the quality of the control action applied to the system at that moment in time indicating how well the action helped achieve the goal or optimize the system's performance.
The instantaneous reward, i.e., reward at a moment in time, is calculated by designing an appropriate reward function that maps the state-action space into the set of real numbers $\mathbb{R}$. The choice of an appropriate reward function depends on how effectively it can discriminate the changes happening in the environment when a control action is applied. As explained in the Subsection~\ref{subsec:Problem_statement}, our aim is to maximize the performance under a given PCAP. By negating the power consumption measured at time $t$ during each of the sampling instance, we can easily take care of the power reduction i.e., $R(s,a) = - measured \ power (t)$. This reward function penalizes an action that can lead to an increased power consumption, but at the same time ignores its effect on performance. 

An alternate choice could be $R(s,a) = - measured \ power(t) \ + \ progress(t)$, however, this is not a good choice for a reward function since it fails to record the trends in the power and performance. For example: A higher progress at a lower PCAP cannot be differentiated from another that may return a lower progress at a higher PCAP. Therefore, a suitable choice of reward function that can differentiate these modes of power and performance is $R(s,a) = \frac{progress}{measured \ power}$.   

To give extra emphasis on the reduction of power, we decided to consider the following reward function consisting of a linear combination of the total measured power and the instantaneous progress to PCAP ratio:
\begin{equation}
                \begin{aligned}
                            R(t) = \underbrace{-c_1* PCAP}_{\text{minimize consumed power}} + \underbrace{c_2 * \frac{progress(t)}{measured \ power (t)}}_{\text{maximize performance per watt}}
                \end{aligned}
                \label{eq:reward_function}
            \end{equation}
where $c_1$ and $c_2$ are scaling factors which we will determine by observing a large number of training performances. In the mathematical model developed using the static characterization in Equation~\ref{non_linear_equivalent}, the measured power is computed as $measured \ power = e^{-\alpha(a.PCAP+b-\beta)}$ which we will be using for the learning process.  
We tested our algorithm with a variety of values for $c_1$ and $c_2$ and observed that more than one value exists as a valid candidate for $c_1$ and $c_2$. One interesting combination which we used for the training and testing is $c_1 = 0$ and $c_2 = 4.44$ which focused only on the maximization of the performance under the given power cap. Therefore, under the present simulation conditions, a reward function of $R(s,a) = \frac{progress(t)}{measured \ power (t)}$ is expected to be equally good when compared to an enhanced power reducing reward function given by Equation~\ref{eq:reward_function}.

\subsection{RL Agent Training}\label{subsec:Training}
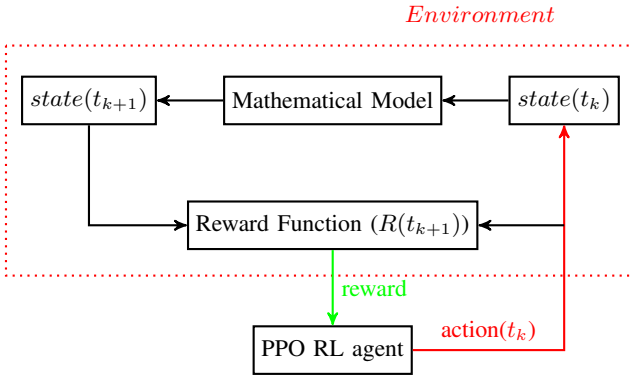
\begin{figure}[htb]
\vspace{-0.5cm}
\begin{tikzpicture}[scale=1, auto, >=stealth']
\centering
\small
\matrix[ampersand replacement=\&, row sep=0.2cm, column sep=0.4cm] {
\node[block](next_state){$state(t_{k+1})$}; \& 
\node[block](model){Mathematical Model}; \& 
\node[block](state){$state(t_k)$};
\\
\\
\\
\\
\\
\&\node[block](reward){Reward Function ($R(t_{k+1})$)};  \& \node[guide](tap){};
\\
\\
\\
\\
\\
\&  \node[block](agent){PPO RL agent};
\\
\\
};
\draw[red,thick,dotted] ($(next_state.north west)+(-0.2,0.4)$) rectangle  ($(state.south east)+(0.2,-2)$);
\node[above] at ($(model.north east)+(0.5,0.6)$){$\color{red}Environment$};
\draw [connector, thick, red] (agent) -| node[above, near start]{action($t_k$)}(state);
\draw [connector] (state) -- node{}(model);
\draw [connector] (model) -- node{}(next_state);
\draw [connector] (next_state) |- node{}(reward);
\draw [connector, thick, green] (reward) -- node[right]{reward}(agent);
\draw [connector] (tap.center) -- node{}(reward);
\end{tikzpicture}
\vspace{-0.5cm}
\caption{Training of the RL-agent as described in the paper. The RL-agent receives a reward, for every action it takes, calculated using the values of state variables obtained from the mathematical model.}
\label{fig:flow_chart_learning}
\end{figure}

The RL system is trained using the mathematical model given by the Equation~\ref{eq:linear_mathematical_model} which is linear in nature. Here, we follow a model-based training approach that is used in a variety of applications~\cite{williams2017information}. A model-based training has the advantage of not risking the safety and stability of hardware while evaluating a variety of control actions. Therefore, while training, we used the entropy term of the PPO agent to enable exploration.
This helped in learning an optimal policy faster. The flow diagram in Figure~\ref{fig:flow_chart_learning} 
shows the general idea employed under the training using the mathematical model.

\section{Empirical Evaluation of Reinforcement Learning Approach}
\label{sec:Eval}

This section presents details of evaluating our approach for its efficacy in meeting its objectives.
First we present the criteria using which we evaluate the proposed method for controlling a data center compute node. 
We then present the results and our analysis.

\subsection{Evaluation Strategy} \label{subsec:evaluation_steps}
We compare here the total time taken for executing the workload and the total energy consumption during that period, with two standard operating conditions of a data center compute node. The evaluation consists of the following analysis based on which we draw conclusions about the efficiency (percentage value of increase or decrease in the execution time and power consumption) of the proposed method: 
\begin{itemize}
\item A comparative analysis of the proposed RL-based control method with the PI controller proposed in prior work~\cite{cerf2021sustaining},
\item A comparative analysis of the RL-based control with the efficiency of the system with minimum and maximum power caps,
\item A repeatability analysis of the experiment on the same compute node, and
\item A repeatability analysis involving different compute nodes.
\end{itemize}
For the comparative analysis with the PI controller, we test our RL agents that are trained using a variety of reward functions and generated using varying values of $c_1$ and $c_2$ while running the benchmark application. Similarly we use the optimal controller among these trained models to control the PCAP, while running the benchmark application, and compare it against the maximum and minimum PCAP runs. We repeat this experiment 10 different times on a node to analyze the statistics associated with the results. We also test the algorithm on different nodes and analyze the statistics for the repeatability of the results on varying hardware.
Through these evaluations, we also comment on the execution time, total power consumed, and the related statistics for each experiment. 

\subsection{Implementation of RL Agent and Experimental Platform}\label{sec:platform}
The code for the training and testing the algorithms are written in Python3 with the required support packages installed on the Nix-environment manager~\cite{dolstra2008nixos}. ~\cite{raffin2021stable} introduces a Python3-based software package called stable-baselines-3 (SB-3) that give users a leverage to code RL algorithms seamlessly. With the support of PyTorch~\cite{paszke2019pytorch}, we developed an easy to implement architecture for all the standard RL methods shown in the literature.

The RL agent resides inside the the Argo Node Resource Manager (NRM) stack~\cite{anrm}.
Argo NRM is an infrastructure developed as part of the U.S. Department of Energy Exascale Computing Project called Argo, for the design of node-level resource management policies. It provides users an easy-to-use and a unified interface (through Unix domain sockets), to the various monitoring and resource control knobs available on a compute node (e.g., RAPL, performance counters). 

All experiments in this paper were performed on Skylake compute nodes using the STREAM benchmark.  
STREAM~\cite{mccalpin1997survey} is a standard application benchmark comprising a set of simple, portable programs that can be used to measure memory bandwidth on a variety of computer systems. The benchmark measures the memory bandwidth for four different types of operations: copy, scale, add, and triad. STREAM is chosen as it is representative of memory-bound phases of applications and shows a stable behavior. STREAM is also easy to modify into an iterative application: its four kernels are ran a configurable number of times in a loop, with a heartbeat being reported to the NRM each time the loop completes, i.e., after one run of the four kernels. 

All four tests are performed on arrays of floating-point numbers and the results are reported in terms of the achieved memory bandwidth (in bytes/second).  The problem size is set to 33,554,432 with 10,000 iterations for the experiment during its evaluation.
In the context of the STREAM benchmark, the problem size refers to the amount of data in bytes that will be used to execute the benchmark. The problem size can be varied to measure the memory bandwidth at different data set sizes. The iterations refer to the number of times that the benchmark will be executed for a given problem size, and the results of the iterations are averaged to reduce the impact of measurement variability. Typically, a large number of iterations are used to obtain accurate and consistent results. 

Note that availability of RAPL actuators and sensors on the compute nodes are important for our experiments. Moreover, each node should have a reasonable range of operation region under the given power cap before it attains saturation so that a number of them can execute the STREAM benchmark seamlessly. We use Chameleon Cloud~\cite{keahey2020lessons}, a server facility made available for researchers to perform experiments that requires low-level access to hardware. Hosted on a variety of locations spanning across the country, it offers hardware components of various specifications. We use the Skylake nodes from the available servers for our experiments that use an Intel Xeon Gold 6126 processor with 12 dual threaded cores. Above all, it belongs to the modern series of the Intel processors with RAPL actuators and sensors.

\subsection{Experiment Details}\label{sec:exp_details}
\begin{figure}[htb]
    \centering
    \includegraphics[trim={0.5cm 0.1cm 0cm 0.5cm},clip,width = \linewidth]{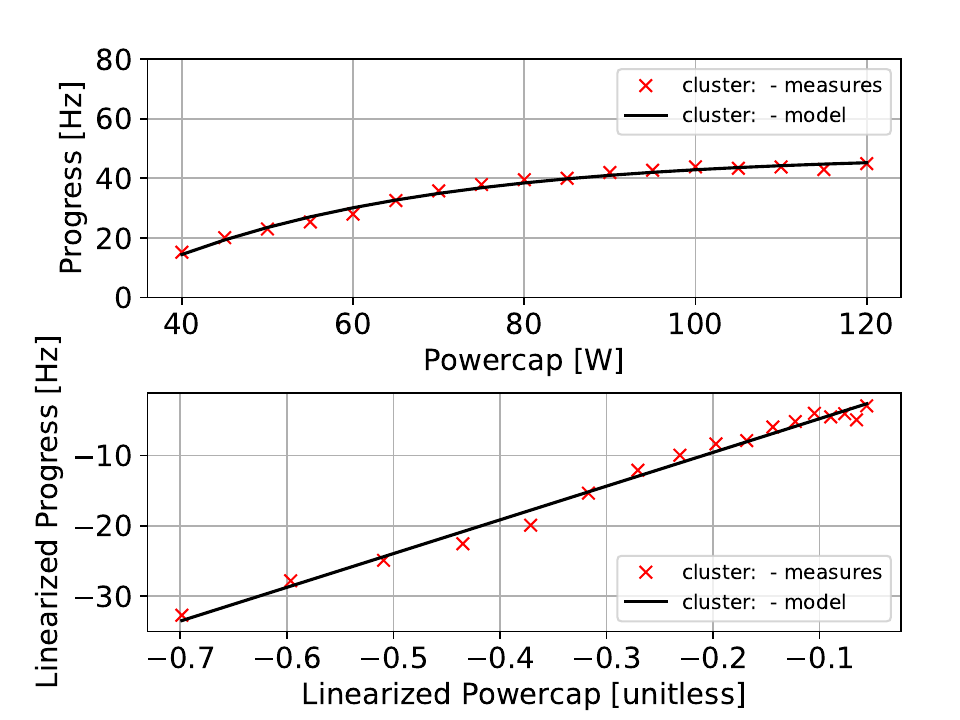}
    \vspace{-0.5cm}
    \caption{Mathematical model of an HPC node representing the non-linear and the corresponding linear relations existing between PCAP and progress, obtained using static characterization, showing the data points (red '$\times$') and the best fit curve (black '$-$').}
    \label{fig:static_characterization}
\end{figure}

For the first part of the experiment involving the static characterization, described in the Subsection \ref{subsec:Mathematical_Modeling}, we used a Chameleon Cloud node to run the STREAM benchmark while the daemon collected the progress measurement. We actuated the RAPL power knobs with test PCAP values in the form of step signals. A total of 17 step signals were given and the progress values were calculated using Equation~\ref{eq:progress-calculation} and recorded once the steady state was attained. We then collected these datasets consisting of the PCAP-progress pairs and used the Scipy line fitting algorithm to obtain the best fit equation. Figure~\ref{fig:static_characterization} depicts the non-linear and linear representations of data points and the curve that was obtained using static characterization, respectively. The parameters obtained after the line fitting are given in Table~\ref{tab:static-coefficients}.


\begin{table}[htb]
\setlength{\tabcolsep}{1em}
\begin{center}
\begin{tabular}{*{5}{l}}
	\toprule
	Description & Notation & Unit & Value \\
	\midrule
	RAPL slope & {$a$} & [1] & 0.95 \\
	RAPL offset & {$b$} & [W] & 0.15 \\
	            & {$\alpha$} & [W\textsuperscript{-1}] & 0.041\\
	power offset & {$\beta$} & [W] & 24.3  \\
	linear gain & {$K_L$} & [Hz] & 47.9  \\
	time constant & {$\tau$} & [s] & 1/3 \\
	\bottomrule
\end{tabular}
\end{center}
\caption{Model parameters obtained using static characterization of an Intel Xeon Gold 6126 processor.}
\label{tab:static-coefficients}
\vspace{-0.5cm}
\end{table}


In the next part  of the experiment, the obtained parameters were used in the mathematical model shown in Equation~\ref{non_linear_equivalent} for generating the next state (progress) in response to an action (PCAP). We employed a PPO-RL agent, to generate a policy and execute it on the mathematical model thereby generating the next state followed by the corresponding reward given by Equation~\ref{eq:reward_function}. At this point, we would like to bring to the reader's attention that we are not running the application nor the NRM daemon while using the mathematical model. We can implement this training algorithm on any standard processor and do not require any specific hardware. Therefore a simple Python code deploying the PPO-RL agent evaluated the policy, followed by periodic updates which assured its convergence to an optimal value. 

For determining the best reward function that could generate an optimal policy, we performed the training using different reward functions, obtained by varying the $c_1$ and $c_2$ in the Equation~\ref{eq:reward_function}. We then collected all the models and tested it on the mathematical model to determine its impact on the theoretical performance and power, and determined the region of interest. We also tested these models on the hardware running the application and daemon to compare it against the PI controller.

Having determined the optimal reward function, we used the optimal model on the hardware for testing followed by evaluation.
For the testing and evaluation, we reintroduced the hardware and the workload to execute, with the help of NRM. This time, the trained RL-agent provided the PCAP for controlling the power knob of the RAPL actuators. Following this, the corresponding progress values were measured  and fed back into to the RL network to compute the next PCAP. This cycle is repeated till the entire work load is finished execution. 
We repeated the experiment for a variety of scenarios described in the Subsection\ref{subsec:evaluation_steps}.

\subsection{Empirical Results and Discussion} 
\label{Sec:Results}
In this section we will explain the results obtained after following the experiment design described in the previous section. We will also perform the evaluation of the algorithm followed by the interpretation of the results obtained.
  
Firstly for choosing an appropriate reward function we trained the RL-agent using the reward function given by the equation \eqref{eq:reward_function} with varying values of $c_1$ and $c_2$ followed by testing on the mathematical model for its performance.
 The theoretical values of performance (execution time) and the total energy consumed during the execution, were obtained using the control generated by each of the agents and are as shown in the Figure \ref{fig:execution_time_vs_power}.
 By using a series of values for $c_1$ and $c_2$ starting from 0 followed by increments of 0.1, till 10, we were able to observe the variations in the performance and power for each of the designed reward. As expected we obtained a combination of values for $c_1$ and $c_2$ that yielded respective results ranging from the fastest to the slowest performance as shown in the figures \ref{fig:execution_time_vs_power}.
Observing the figure we can see that there is a point in the maximum curvature region (marked in blue circle) in figure \ref{fig:execution_time_vs_power} that can be considered as an ideal region to look for a candidate reward function. This point corresponds to the minimum energy consumption and the maximum performance. Note that, there were a couple of ideal candidates for the reward function that repeated across multiple execution of this experiment, for example, ($c_1$=1.052,$c_2$=2.22) and ($c_1$=0,$c_2$=4.44) gave the same responses. We chose the model, that was obtained after training using the reward function, $R(t) = -1.052 PCAP(t) + 2.22 \frac{progress(t)}{measured \ power(t)}$, as the RL-agent to control the HPC nodes during the testing phase.
\def\sca{1}
\def\scb{1}
\begin{figure}[htb]
  \centering
  \begin{tikzpicture}[remember picture]
    \node[inner sep=0pt,anchor=south west] (A1) {\includegraphics[trim={0.9cm 0cm 1cm 1cm},clip,width=\sca\linewidth]{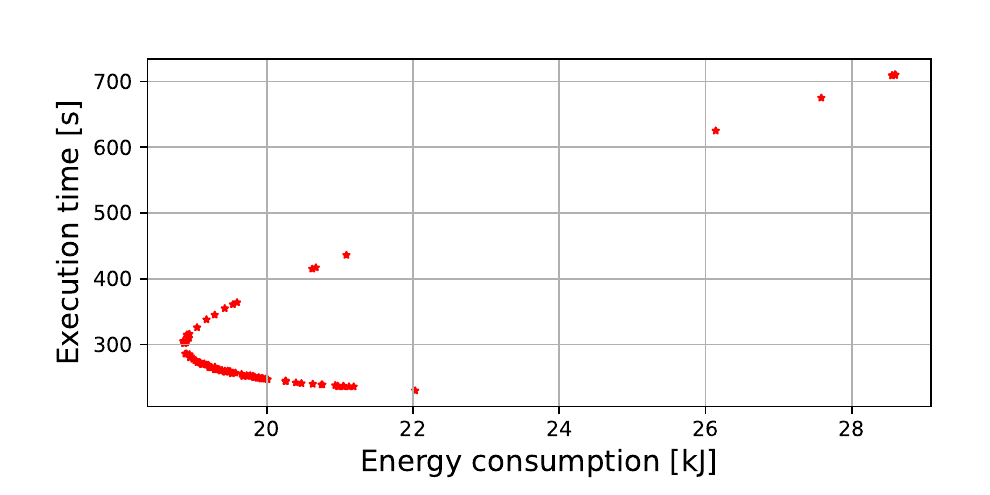}};
    \draw[line width=0.4mm,blue] (\sca*1.56,\sca*1.4) circle (10pt) node[anchor=west]{};
  \end{tikzpicture}
  \begin{tikzpicture}[remember picture,overlay]
  \end{tikzpicture}
  \vspace{-0.75cm}
  \caption{A Pareto of execution time vs energy consumed, generated by varying the RL-agents trained with different reward functions. The x-axis shows the total energy consumed during the execution, and the time of execution is plotted on the y-axis. The ideal characteristics are inside the blue circle.}
  \label{fig:execution_time_vs_power}
\end{figure}

\subsection{Analysis of Results}\label{subsec:Evaluation}
\begin{figure}[htb]
    \centering
    \includegraphics[trim={0.5cm 0.4cm 1.5cm 1.9cm},clip,width = \linewidth]{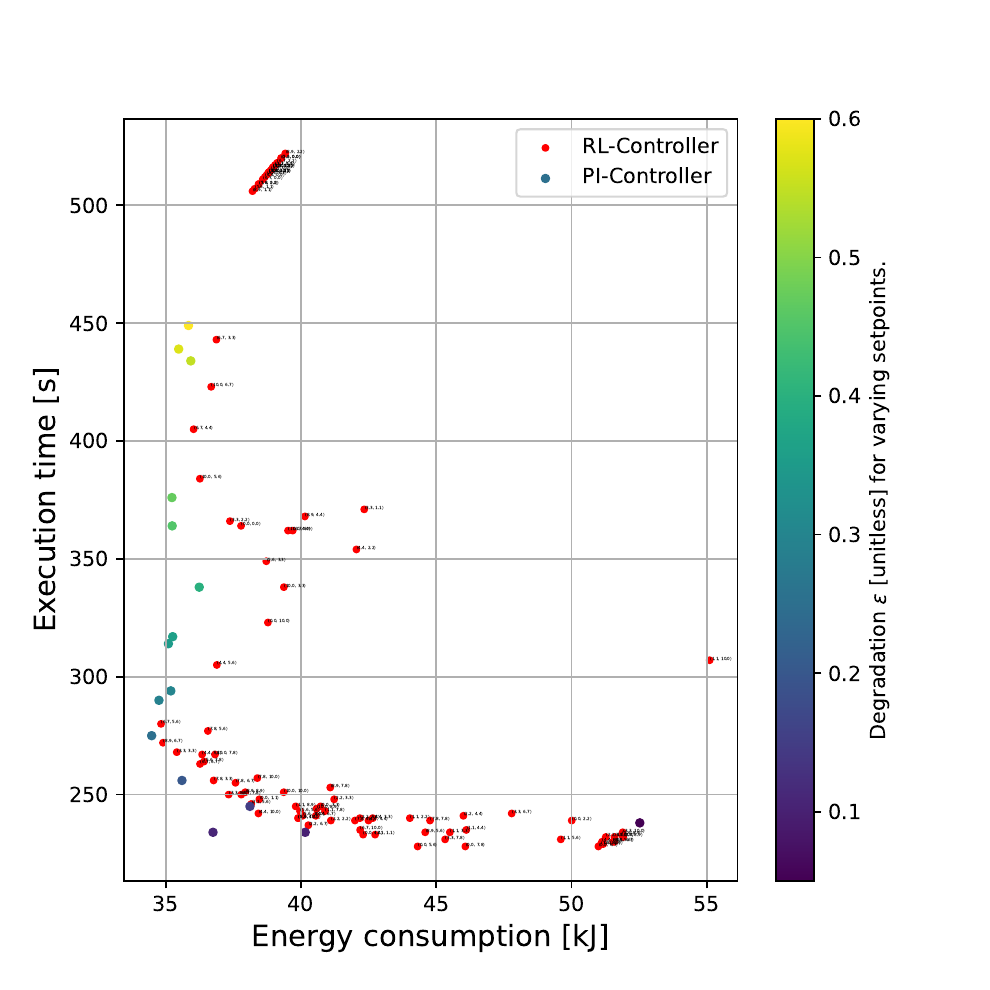}
    \vspace{-0.5cm}
    \caption{Plot showing the comparison of PI controller with varying setpoints (values for $1- \epsilon$) and RL controller for varying reward functions. The PI control objective is given as a degradation factor $\epsilon$, that is, the tolerable loss of performance}
    \label{fig:comparison_PI_RL_controller}
\end{figure}

Before we proceeded with the analysis and evaluation of the proposed optimal controller for the PCAP, we needed to verify its performance by comparing it against the existing PI controller. For that, we needed to run both the algorithms on the same machine. We first executed the PI controller on the HPC node with varying set-points as shown in the Figure ~\ref{fig:comparison_PI_RL_controller}. The pareto of points with gradients of colors ranging from yellow to blue depicts the result obtained for the PI controller with varying setpoints. The setpoint was specified using the parameter $\epsilon$, representing the allowed tolerance in the maximum performance value. The $\epsilon$ value was varied from 0 to 0.6 randomly, during the experiments using PI controller. The shortest execution time of 240\unit{\second} (corresponding to faster execution) was obtained at an energy consumption of 36\unit{\kilo \joule} at an $\epsilon$ value of 0.1.
The setpoint value is expected to change based on the hardware being tested, whose dependency on the controller is being removed while using an RL controller. 
We then tested the trained RL controllers on the same hardware running the application and created a similar pareto of points. In the Figure ~\ref{fig:comparison_PI_RL_controller}, the red points represent the response of the HPC node controlled using the trained RL-agent. We observed that the optimal region consisted of points vary close to the best performance obtained using the PI controller. It was observed that the $R(t)$ proposed in the previous subsection fell within a closer range of the best performance.

Through these comparisons we were able to justify the use of a trained RL-agent as an optimal controller aimed to minimize the power consumption. Using the proposed RL controller, we were able to regulate the HPC node performance around those regions of operation, where a PI controller was regulating the performance, given a user-defined set-point. We also were able to excite the HPC node at some other regions where a slower response was obtained but at a very low energy consumption. We will now look for points where optimal performance and power consumption was recorded and will check for the repeatability of the experiments using the same hardware as well as different hardware.

We used the RL-agent trained using the reward function chosen previously, for the remaining part of the experiment consisting of:
\begin{itemize}
\item Repeatability on the same node and
\item Repeatability on different nodes.
\end{itemize}

Within each of these analysis, we considered two different scenarios to compare with the performance of the proposed method, namely:
\begin{itemize}
\item Maximum PCAP: allowing the node to utilize the maximum allowed capabilities of the system,
\item Minimum PCAP: limiting the performance, by allowing only the minimum power at the actuators.
\end{itemize}

\begin{figure}[htb]
    \centering
    \includegraphics[trim={0.5cm 0.5cm 0.5cm 2cm},clip,width=\linewidth]{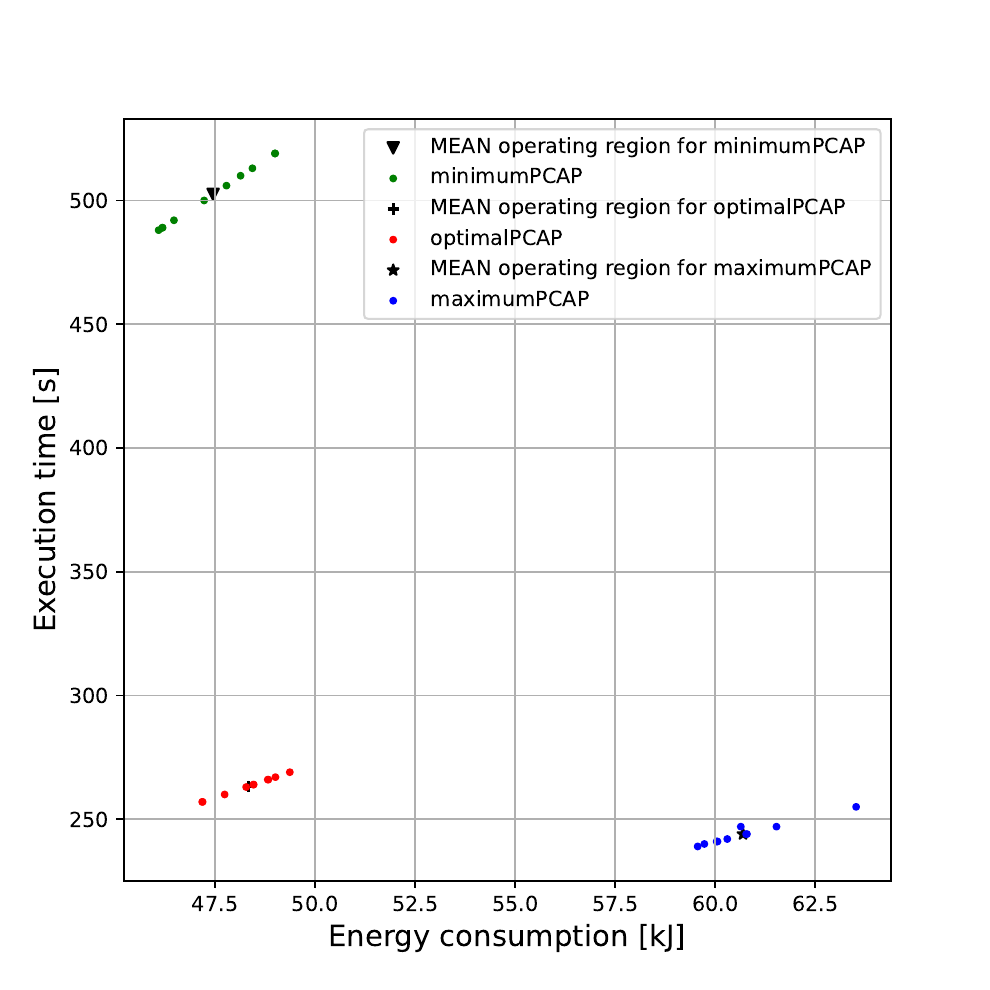}
    \caption{Comparison of the execution-time and the energy consumed during the execution of the experiments performed on the same hardware employing three different methods. The statistics of the executions are shown in the table \ref{tab:statistics_repeatability}}
    \label{fig:comparison_execution_time}
\end{figure}
\begin{figure}[htb]
    \centering
    \includegraphics[trim={0.5cm 0.5cm 0.5cm 2cm},clip,width=\linewidth]{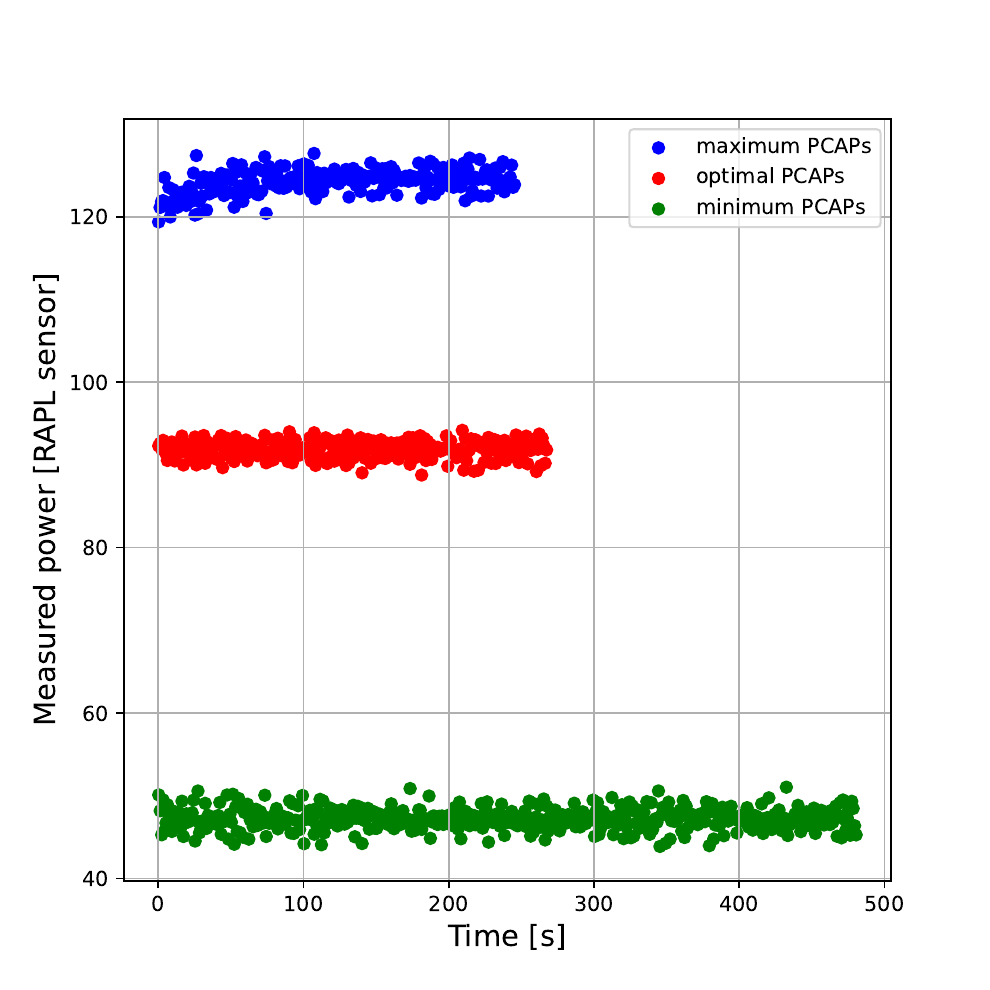}
    \caption{Instantaneous power vs time-steps plots for a single execution of the experiment using optimal, maximum and minimum PCAP controllers.}
    \label{fig:comparison_power}
\end{figure}

For the analysis of the repeatability of the experiment on the same node, we conducted a total of 30 executions of the experiment with RL controller being used for 10 of the executions. In the remaining 20 executions we used the maximum allowed and minimum allowed values for the PCAP to record the statistics. We didn't consider the PI controller for the comparison here, since we are only evaluating the repeatability of the proposed method. We then observed the statistics associated with each set of executions and the values are tabulated.  
Figure \ref{fig:comparison_execution_time} depicts the results of the experiment with the execution time plotted against the energy consumed during the each of the experiments. Figure \ref{fig:comparison_power} shows the values of instantaneous PCAPs sensed during each time step, by the RAPL sensors for three different executions using maximum, minimum and optimal PCAP respectively. 
{
In Figure \ref{fig:comparison_execution_time}, ensuring the maximum PCAP for the RAPL actuators, that put no limit on the amount of energy drawn into the system, the workload execution was completed in 245\unit{\second}. But when we look at the total energy consumed over the period it was in the range of an average value of 60.97\unit{\kilo\joule}. On the other hand, enforcing the minimum PCAP on RAPL actuators, the power consumption was minimal, i.e in the range of an average value of 47.49\unit{\kilo\joule} of energy, but compromising on the total execution time which lasted for 501.14\unit{\second}. On the other hand, the RL based controller was able to limit the power consumption and comparable execution time. It was observed that the average power consumption, over 10 experiments using the optimal PCAP, was 48.23\unit{\kilo\joule}, while the average execution time was 261.19\unit{\second}. This shows a reduction of 21\% on the consumed power, while the performance was compromised only by 6.5\%.} The statistics of the experiment are given in the table \ref{tab:statistics_repeatability}
\begin{table}[htb]
\setlength{\tabcolsep}{1em}
\begin{center}
\begin{tabular}{*{5}{l}}
	\toprule
	{} & Minimal & Maximal & Optimal \\
	\midrule
	$\mu$ Execution Time [\unit{\second}] & 501.14 & 245.23 & 261.19.\\
    $\sigma$ Execution Time & 26.20.21 & 3.6 & 5.89 \\
    $\mu$ Energy Consumption [\unit{\kilo \joule}] & 47.49 & 60.97 & 48.23 \\
    $\sigma$ Energy Consumption & 1.98 & 0.77 & 0.744 \\
	\bottomrule
\end{tabular}
\end{center}
\caption{Statistics of repeatability using the same node over 10 executions. $\mu$ and $\sigma$ depict mean and standard deviation respectively.}
\label{tab:statistics_repeatability}
\vspace{-0.5cm}
\end{table}

For the next part, we performed a total of 21 exections consisting of 7 sets with 3 executions each with maximum, minimum and optimal PCAP. Each set was exectuted on different hardware. We made the reservation for 7 nodes from the Chameleon Cloud using a single lease in-order to assure diversity in hardware. The results of the execution was then gathered for analysis of its statistics. 
\begin{figure}[htb]
    \centering
    \includegraphics[trim={0.5cm 0.5cm 0.5cm 2cm},clip,width=\linewidth]{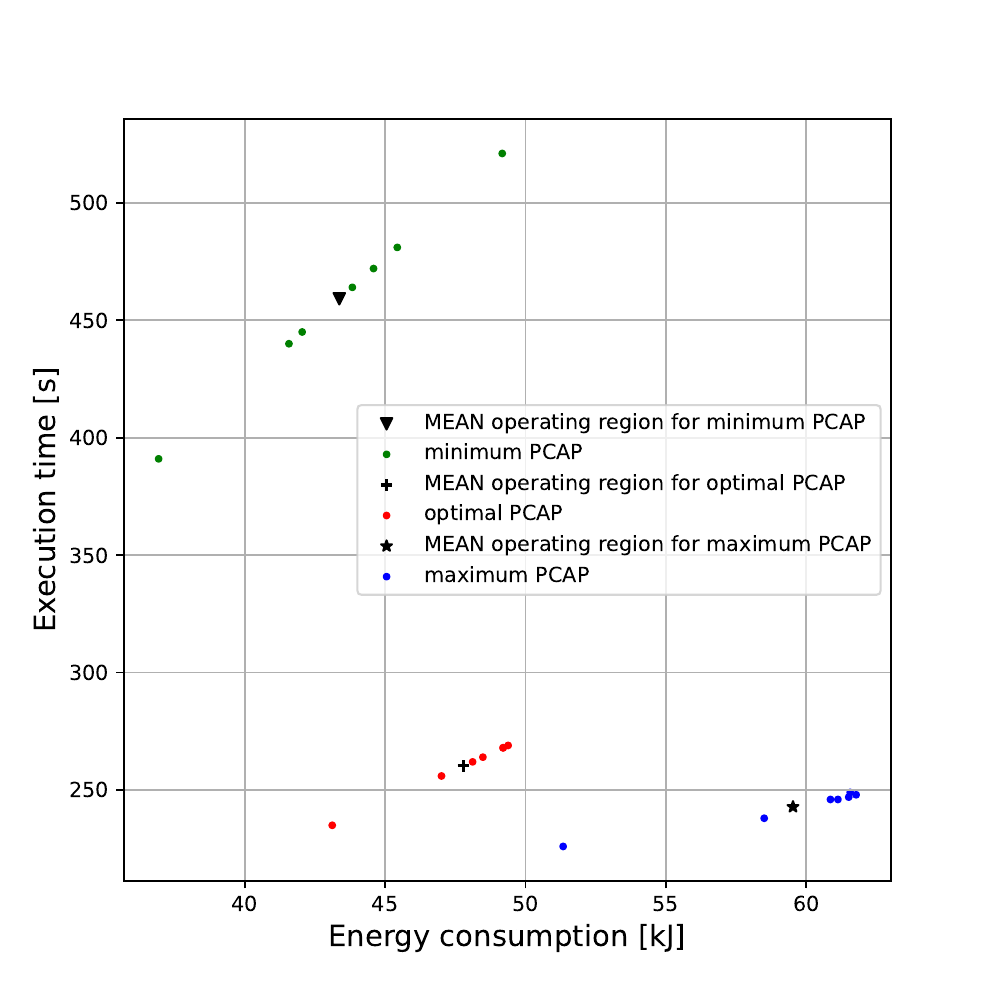}
    \caption{Comparison of the execution-time and the energy consumed during the execution of experiments, performed on different nodes using the three methods. The statistics of the executions are shown in the table \ref{tab:statistics_repeatability_diff_nodes}}
    \label{fig:comparison_execution_time_diff_nodes}
\end{figure}
Figure \ref{fig:comparison_execution_time_diff_nodes} shows the obtained results of the experiment. It was observed that the proposed RL based controller was able to bring down the energy consumption by 19.7\% at a cost of 7.3\% in performance.
It was also observed that the variance among each set of the experiments have gone high due to the changes in hardware, but in effect the optimal controller performed with the minimum variance. The related statistics are presented in the table \ref{tab:statistics_repeatability_diff_nodes}.
\begin{table}[htb]
\setlength{\tabcolsep}{1em}
\begin{center}
\begin{tabular}{*{5}{l}}
	\toprule
	{} & Minimal & Maximal & Optimal \\
	\midrule
	$\mu$ Execution Time [\unit{\second}] & 459.14 & 242.85 & 260.28\\
    $\sigma$ Execution Time  & 40.21 & 8.25 & 12.03 \\
    $\mu$ Energy Consumption [\unit{\kilo \joule}] & 43.3 & 59.52 & 47.78 \\
    $\sigma$ Energy Consumption & 3.78 & 3.77 & 2.22 \\
	\bottomrule
\end{tabular}
\end{center}
\caption{Statistics of repeatability over 5 executions on different HPC nodes. $\mu$ and $\sigma$ have the usual meaning.}
\label{tab:statistics_repeatability_diff_nodes}
\end{table}
Through the repeatability experiments we were able to prove the reliability of the proposed approach for generating an optimal control for HPC nodes. We were also able to verify that the energy consumption reduction is consistent throughout its execution irrespective of the hardware.

\balance
\section{Conclusion}\label{sec:Conclusion}
This paper addresses the problem of managing power consumption in heterogeneous data center compute nodes by focusing on the potential of dynamically adjusting power across
compute elements to save energy with almost no impact on performance.
Our approach involves model-based reinforcement learning that learns using a mathematical model, which is generated using static characterization of a specific hardware node running a standard application benchmark. The control of the hardware, then uses the trained RL agent to ensure optimal performance. Experimental validation shows promising results for systems running a memory-bound benchmark. The repeatability of results showed a standard deviation of 5.89 with a mean energy consumption of 48.23 \unit{\kilo \joule} during 261.19\unit{\second} of mean execution time. These results indicate that the RL-based approach can serve as an alternative to the PI controllers used in our past experiments in order to remove the dependency on operational setpoints. The approach presented in this paper is most suitable for cloud providers and high performance compute node administrators who seek solutions to reduce power consumption of their server resources yet not impact application performance. In our future work, we aim to resolve the dependency of the proposed work on the mathematical model for training. We also seek to design a generalized RL-based controller, controlling PCAPs on nodes executing a variety of applications.

\section*{Acknowledgments}\label{sec:Acknowledgments}
Results presented in this paper were obtained using the Chameleon testbed supported by the National Science Foundation.
Argonne National Laboratory's work was supported by the U.S. Department of Energy, Office of Science, Advanced Scientific Computer Research, under Contract DE-AC02-06CH11357. This research was supported by the Exascale Computing Project (17-SC-20-SC), a collaborative effort of the U.S. Department of Energy Office of Science and the National Nuclear Security Administration. The dataset and the codes used for the entire experiment, including the static characterization, training and testing, are hosted on our  \href{https://github.com/akhileshraj91/GENERALIZED_RL_ANL.git}{GitHub repository}. 
We also provide readers with complete access to our \href{https://chi.tacc.chameleoncloud.org/ngdetails/OS::Glance::Image/f6974b54-93fc-4140-9163-7aae9fc3070d}{Chameleon Cloud image}, which will enable them to reproduce the results.

\balance
\bibliographystyle{ieeetr}
\bibliography{refs.bib}
\end{document}